\documentstyle[psfig,12pt]{article}

 \newfont{\titlefont}{cmssbx10 scaled\magstep5}

\newcommand{\CD}{{\cal D}}

\newcommand{\CL}{{\cal L}}

\newcommand{\CO}{{\cal O}}

\newcommand{\CZ}{{\cal Z}}

\newcommand{\bea}{\begin{eqnarray}}  \newcommand{\eea}{\end{eqnarray}}
\newcommand{\beq}{\begin{equation}}  \newcommand{\eeq}{\end{equation}}
\newcommand{\non}{\nonumber}  
  
\newcommand{\lmk}{\left(}  \newcommand{\rmk}{\right)}
\newcommand{\lkk}{\left[}  \newcommand{\rkk}{\right]}
\newcommand{\lhk}{\left \{ }  \newcommand{\rhk}{\right \} }
\newcommand{\del}{\partial}  
\newcommand{\vect}[1]{\mbox{\boldmath${#1}$}}
\newcommand{\bib}{\bibitem} \newcommand{\new}{\newblock}
\newcommand{\la}{\left\langle} \newcommand{\ra}{\right\rangle}
 
\newcommand{\gtilde} {~ \raisebox{-1ex}{$\stackrel{\textstyle >}{\sim}$} ~} 
\newcommand{\ltilde} {~ \raisebox{-1ex}{$\stackrel{\textstyle <}{\sim}$} ~}

\newcommand{\eff}{\rm eff}

\begin{document}

\begin{flushright}
  UTAP-281 \\ YITP-97-55  
\end{flushright}

\begin{center}
  {\Large \bf Probability Distribution Function of the Coarse-grained 
   Scalar Field at Finite Temperature \\}
  \vskip 1cm
  {\large Masahide Yamaguchi} \\
  \vskip 0.2cm
  {\large \em Department of Physics, University of Tokyo} \\
  \vskip 0.1cm
  {\large \em Tokyo 113, Japan}
  \vskip 0.5cm {\large Jun'ichi Yokoyama} \\ 
  \vskip 0.2cm {\large \em Yukawa Institute
    for Theoretical Physics, Kyoto University} \\ 
  \vskip 0.1cm {\large \em Kyoto 606-01, Japan}
  \vskip 0.5cm
 
\end{center}


\begin{abstract}
We present a formalism to calculate the probability distribution
function of a scalar field coarse-grained over some spatial
scales with a Gaussian filter at finite temperature. As an
application, we investigate the role of subcritical fluctuations in the
electroweak phase transition in the minimal standard model.
It is  concluded that the universe was in a
mixed state of true and false vacua already at the critical temperature.
\end{abstract}

\thispagestyle{empty} \setcounter{page}{0} \newpage
\setcounter{page}{1}


\section{Introduction}

\label{sec:introduction}

\indent

Quantum field theories with spontaneous symmetry breaking usually
exhibit symmetry restoration at high temperature \cite{KL}, and hence
our universe is supposed to have experienced several phase transitions
in the early stage of its evolution as the cosmic temperature
decreased with expansion.  Much work has been done on these phase
transitions and their generic properties such as the transition order
and the critical temperature, $T_c$, have been extracted from the
effective potential.  The dynamics of phase transitions, on the other
hand, is less understood, partly because it is difficult to clarify
the role of thermal fluctuations.

Thermal fluctuations often play a conspicuous role on the dynamics of
phase transitions. For example, at formation of topological defects
associated with some symmetry breaking \cite{Kib}, magnitude of
thermal fluctuations fixes the temperature when the phase distributed
randomly on each coherent volume becomes stabilized. Also, their
statistical property is important to set the initial distribution of
topological defects. Furthermore,  thermal fluctuations
called subcritical bubbles have been considered  important in the context
of the electroweak phase transition, where the one-loop improved
finite-temperature effective potential shows that it is of first order
\cite{DLHLL}-\cite{She}. Gleiser, Kolb, and Watkins \cite{GKW}
\cite{GK}, however, suggested that the barrier between the two minima
of the effective potential at $T_c$ was so shallow that thermal
fluctuations called subcritical bubbles might be dominant and that the
standard bubble nucleation picture of the first order phase transition
might not work.

Thus, it is indispensable to understand the property of thermal
fluctuations in order to elucidate the real dynamics of the phase
transition. Among the property of thermal fluctuations, their typical
scale and amplitude are particularly important. The former gives us
the scale we should focus on, and the latter determines whether
thermal fluctuations dominate the dynamics. First, we should notice
that in evaluating the magnitude of thermal fluctuations, we should
pay attention to fluctuations coarse-grained over some characteristic
coherent scale of thermal fluctuations rather than raw fluctuations.
If we estimate the typical amplitude out of raw fluctuations, random
infinitesimal ones make a dominant contribution to them and we cannot
obtain a meaningful value to understand the role of thermal
fluctuations.

In this paper we present a formalism to calculate the
probability distribution function (PDF) of a coarse-grained scalar
field at finite temperature. The PDF of coarse-grained fluctuation has
also been studied by Hindmarsh and Rivers \cite{HR} and by Bettencourt
\cite{Bet}, but they gave only the upper limit of probability.

As an application, we apply our formalism to the electroweak phase
transition which may be cosmologically important for the possibility
of baryogenesis during it \cite{KRS}. For successful baryogenesis, it
should be of first order to realize an out-of-equilibrium condition
\cite{Sak}. However, if subcritical fluctuations dominate the dynamics
of the phase transition it proceeds in a similar manner to a
second-order transition and baryogenesis becomes impossible.  Here we
evaluate PDF of the Higgs field on the relevant scales in the minimal
standard model and investigate whether the universe is homogeneously
in the symmetric phase at the critical temperature.  We have also
studied this issue in the previous paper \cite{YY}, solving a
phenomenological Langevin equation for the {\it classical} expectation
value of the Higgs field numerically.  But the PDF we formulate here
is based on finite-temperature field theory and quantum effects are
taken into account more appropriately.

In the next section, we present a formalism of evaluating the PDF for
the coarse-grained fluctuations. In Sec.\ \ref{sec:application}, the
method established in Sec.\ \ref{sec:formalism} is applied to the
electroweak phase transition.

\section{Formulation of the probability distribution function}

\label{sec:formalism}

\indent

We start with the partition function $Z$ of a scalar field
$\varphi(\vect x)$ at the finite temperature $T=\beta^{-1}$, defined
by

\beq
  Z =  \mbox{tr} \rho = 
    \int \CD \varphi(\vect x) 
    \rho_\beta[\varphi(\vect x), \varphi(\vect x)] \:, 
\label{eqn:partition}   
\eeq
 
\noindent
where $\rho_\beta [\varphi(\vect x), \varphi(\vect x)]$ is the
diagonal component of the density matrix, $\rho_\beta$, which gives
the relative probability distribution of the configuration
$\varphi(\vect x)$. Then the PDF of the field, $\tilde P[\varphi(\vect
x)]$, is expressed by path integral as

\bea
 \tilde P[\varphi(\vect x)] 
   &=& \frac{1}{Z} \rho_\beta [\varphi, \varphi] \non \\
   &=& \frac{1}{Z} \int_{B_{1}} \CD \phi(x) 
         \exp \lmk -\int d^4x \CL_{E}[\phi(x)] \rmk \:, 
 \label{eqn:prob}     
\eea

\noindent
where $x = (\vect x, \tau)$, $\CL_{E}[\phi]$ is the Euclidean
Lagrangian density given by

\beq
  \CL_{E}[\phi(x)] = \frac12 (\del\phi)^2 + V[\phi] \:, 
\eeq

\noindent
and the boundary condition $B_{1}$ implies $\mbox{$\phi(\vect x,\tau =
  0)$} = \mbox{$\phi(\vect x,\tau = \beta)$} = \mbox{$\varphi(\vect
  x)$}$.

Now we consider the PDF of the field coarse-grained over an arbitrary
scale $R$ as,

\beq
  \varphi_{R} \equiv \int d^3 x \varphi(\vect x)I(\vect x;R) \:, 
\eeq

\noindent
where $I(\vect x;R)$ is a Gaussian window function with the width $R$
and given by

\beq
  I(\vect x;R) = \frac{1}{(2\pi R^2)^{\frac32}}
                  \exp \lmk -\frac{\vect x^2}{2R^2} \rmk \:. 
\eeq

\noindent
Then the probability that $\varphi_{R}$ is equal to $\bar\varphi$,
$P[\varphi_{R}=\bar\varphi]$, is formally given by

\beq
  P[\varphi_{R}=\bar\varphi] = \int \CD \varphi(\vect x)
           \tilde P[\varphi(\vect x)] \delta(\varphi_{R}-\bar\varphi) \:.
\eeq

\noindent
Inserting (\ref{eqn:prob}) into this equation, we get the following
expression,

\bea
  P[\varphi_{R} = \bar\varphi] 
    &=& \frac{1}{Z} \int \CD \varphi(\vect x) \int_{B_{1}}
          \CD \phi(x) \exp \lmk -\int d^4x \CL_{E}[\phi(x)] \rmk
            \delta(\varphi_{R}-\bar\varphi)   \non \\
    && \hspace{-3.5cm}
        = \frac{1}{Z} \int_{B_{2}} \CD \phi(x)
            \int_{-\infty}^{+\infty} 
            \frac{d\alpha}{2\pi} \exp(-i\alpha \bar\varphi)
              \exp \lmk 
            -\int d^4x \CL_{E}[\phi(x)] + \beta 
              \int d^3 x J(\vect x)\phi(\vect x,0) \rmk \:, 
  \label{eqn:cprob}        
\eea

\noindent
where $J(\vect x) \equiv i\alpha I(\vect x;R)/\beta$ and the boundary
condition $B_{2}$ implies $\mbox{$\phi(\vect x,\tau = 0)$} =
\mbox{$\phi(\vect x,\tau = \beta)$}$.

\noindent
Now, we represent the field, $\phi(\vect x, \tau)$, in terms of the
Matsubara frequency,

\bea
  \phi(\vect x, \tau) &\equiv& \varphi_{0}(\vect x)
      + \sum_{n\ne0}\varphi_{n}(\vect x) e^{2\pi i n \tau/\beta} \non \\
         &\equiv& \varphi_{0}(\vect x) + \varphi_{h}(\vect x,\tau) \:.
\eea

\noindent
Here, $\varphi_{n}$ represents a heavy mode with an effective
mass-squared $m_{n}^2 =V''[\varphi_0] + (\frac{2\pi n}{\beta})^2$ and
$\varphi_{h}$ is the collection of heavy modes. In the high
temperature regime, the heavy modes have large effective masses so
that fluctuations of the heavy modes are much smaller than that of the
zero mode, $\varphi_{0}(\vect x)$, and can safely be integrated out.
For this purpose, the Lagrangian is expanded around the zero-mode,

\bea
  \CL_{E}[\phi(x)] 
    &=& \CL_{E}[\varphi_{0}(\vect x) + \varphi_{h}(\vect x,\tau)]  \non \\
    &=& \CL_{E}[\varphi_{0}(\vect x)] +
         \frac{\delta\CL_{E}}{\delta\varphi_{0}} \varphi_{h}(x) +
          \frac12 \varphi_{h} \frac{\delta^2\CL_{E}}{\delta\varphi_{0}^2} 
           \varphi_{h} +        
            \CL_{E,int}(\varphi_{h};\varphi_{0}) \:.
\eea
        
\noindent
where $\CL_{E,int}(\varphi_{h};\varphi_{0})$ represents higher order
terms in $\varphi_h$ which generate multi-loop effects and will be
omitted hereafter. Here,
$\displaystyle\frac{\delta^2\CL_{E}}{\delta\varphi_{0}^2}$ is given by

\beq
  \frac{\delta^2\CL_{E}}{\delta\varphi_{0}^2} = -\frac{\del^2}{\del
   \tau^2} - \nabla^2 + V''[\varphi_{0}] \:,
\eeq    

\noindent
and for each $n$ mode it acts as

\beq
  \left. \frac{\delta^2\CL_{E}}{\delta\varphi_{0}^2} \right|_{n} = 
   \lmk \frac{2\pi n}{\beta} \rmk^2 -
     \nabla^2 + V''[\varphi_{0}] \:. 
\eeq       

\noindent
Thus the exponent of integrand of the path integral in
(\ref{eqn:cprob}) becomes

\bea
\lefteqn{ -\int d^4x \CL_{E}[\phi(x)] + \beta 
              \int d^3 x J(\vect x)\phi(\vect x,0) = }\\
 \lefteqn{
    -\int d^4 x \lhk 
     \CL_{E}[\varphi_{0}(\vect x)] +
      \frac{\delta\CL_{E}}{\delta\varphi_{0}} \varphi_{h}(x) +
       \frac12 \varphi_{h}(x) \frac{\delta^2\CL_{E}}{\delta\varphi_{0}^2} 
        \varphi_{h}(x) 
          -J(\vect x) 
            \lkk  \varphi_{0}(\vect x) + 
              \sum_{n\ne0}\varphi_{n}(\vect x) \rkk
           \rhk} \non \\
&&  \hspace{-0.5cm} =  \int d^4 x \lhk 
          -\CL_{E}[\varphi_{0}(\vect x)] 
           - \left. \frac12 \sum_{n\ne0} \varphi_{n}^{*}(\vect x)
              \frac{\delta^2\CL_{E}} {\delta\varphi_{0}^2} \right|_{n} 
             \varphi_{n}(\vect x)
          + \lkk\varphi_{0}(\vect x) + 
                   \sum_{n\ne0}\varphi_{n}(\vect x)\rkk J(\vect x)
               \rhk \non \:,   
\eea

\noindent
where $$\int d^4 x \equiv \int_{0}^{\beta} d\tau \int d^3 x \:.$$

\noindent
Note that heavy modes also attach to the external field $J(\vect x)$.
If we first evaluate only the probability distribution functional
$\tilde P[\varphi(\vect x)]$ for the raw field $\varphi(\vect x)$ as
was done by Hindmarsh and Rivers \cite{HR}, the above term is omitted
in integrating out heavy modes. Therefore our formalism is more
rigorous than theirs to this end. Also, as a result, the propagator in
evaluating the connected generating functional $W[J]$, defined below,
is the finite-temperature propagator in Hindmarsh and Rivers
\cite{HR}, while in our formalism it is the zero-temperature
propagator.

After integrating over heavy modes, the PDF
 is given up to one-loop order by

\bea
  P[\varphi_{R} = \bar\varphi] 
   &=& \frac{1}{Z} \int \CD\varphi_{0}(\vect x) \int \frac{d\alpha}{2\pi} 
         \exp(-i\alpha\bar\varphi) \times
    \non \\
   && \hspace{-2.0cm}
        \exp \lkk -\beta H[\varphi_{0}] \
        + \frac{\beta}{2} \int d\vect x 
          \sum_{n\ne0} J(\vect x) \lmk \left. \frac{\delta^2\CL_{E}}
              {\delta\varphi_{0}^2} \right|_{n} \rmk^{-1} 
            J(\vect x)
       + \beta \int d\vect x J(\vect x)\varphi_{0}(\vect x) \rkk \:. \non \\
   &&
\label{eqn:pro3}
\eea

\noindent
where $H[\varphi_{0}]$ is the one-loop finite-temperature
three-dimensional effective {\it action} obtained by integrating over
heavy modes only.  In the case $\varphi$ is coupled to other quantum
fields, integration over these fields in all modes should also be
performed at this stage to obtain the effective action of $\varphi_0$.
 
Here we comment on the difference between the effective action
integrated over only heavy modes and that over all modes. We
concentrate on the homogeneous part, namely the effective {\it
  potential} for simplicity. The effective potential integrated over
all modes is written by the zeta-function as \cite{Ram}

\beq
  F_{\rm all}[\varphi_{0}] = \frac{1}{2\beta U}
           \left. \frac{d\zeta_{\CO}(s)}{ds} 
                  \right|_{s = 0}   \:,
\eeq

\noindent
where $U$ is the box size of the system, or, $\int d^3 x$, and
$\zeta_{\CO}(s)$ is defined by

\beq
  \zeta_{\CO}(s) = \sum_{n} \frac{1}{a_{n}^{s}} \:,
\eeq

\noindent
with $a_{n}$ a series of positive real discrete eigenvalues of an
operator $\CO$. In our case, the operator $\CO$ is identified with
$\displaystyle{\frac{\delta^2\CL_{E}}{\delta\varphi_{0}^2}}$.

\noindent
Then the corresponding zeta-function becomes

\beq
  \zeta_{\CO}(s) = \frac{U}{\Gamma(s)} 
                  \frac{(\mu\beta)^{2s}}{(4\pi\beta^2)^{3/2}}
                   \int_{0}^{\infty} d\sigma \sigma^{s-5/2}
                    \sum_{n = -\infty}^{\infty}
                     e^{-(4 \pi^2 n^2 + V''[\varphi_{0}])\,\sigma} \:,
\eeq

\noindent
where $\mu$ is the renormalization scale and $n$ represents each mode.
Since the zero-mode contribution to $\zeta_{\CO}(s)$ is given by

\beq
  \zeta_{\CO}(s)_{n = 0} = \frac{U V''[\varphi_{0}]^{3/2}}{(8\pi)^{3/2}} 
   \lmk \frac{\mu}{V''[\varphi_{0}]^{1/2}} \rmk^{2s}
      \frac{\Gamma(s-3/2)}{\Gamma(s)}
\:,
\eeq

\noindent
that to $\displaystyle\frac{d\zeta_{\CO}(0)}{ds}$ becomes

\beq
  \left. \frac{d\zeta_{\CO}(0)}{ds}
      \right|_{n = 0} = \frac{U V''[\varphi_{0}]^{3/2}}{6\pi} \:.
\eeq 

\noindent
After all, the difference of the two quantities, namely the zero-mode
contribution to the effective potential, is given by

\beq
  F_{n = 0}[\varphi_{0}] = -\frac{V''[\varphi_{0}]^{\frac32}}{12\pi}T
\:.
\eeq

We now return to the evaluation of eq.\ (\ref{eqn:pro3}). Below we
omit the suffix $0$. We first note the convergency of
$\alpha$-integration in eq.\ (\ref{eqn:pro3}). Using the high
temperature limit, we find

\beq
  \sum_{n\ne0} \lmk \left. \frac{\delta^2\CL_{E}}
      {\delta\varphi_{0}^2} \right|_{n} \rmk^{-1}  
   \simeq \frac{\beta^2}{12} \:.
\label{eqn:high}
\eeq

\noindent
Then $\alpha$-integration becomes

\bea
  && \hspace{-1.5cm} 
    \int_{-\infty}^{\infty} \frac{d\alpha}{2\pi} 
        \exp \lkk -i\alpha\bar\varphi
        + \frac{\beta}{2} \int d^3 x 
          \sum_{n\ne0} J(\vect x) \lmk \left. \frac{\delta^2\CL_{E}}
              {\delta\varphi_{0}^2} \right|_{n} \rmk^{-1} 
            J(\vect x)
       + \beta \int d^3 x J(\vect x)\varphi_{0}(\vect x) \rkk
           \non \\
  && = \int_{-\infty}^{\infty} \frac{d\alpha}{2\pi} 
         \exp \lkk -i\alpha(\bar\varphi - \varphi_{R})
                    - \frac{\beta}{24} \alpha^2
                       \int d^3 x I(\vect x;R)^2 \rkk \non \\
  && = \int_{-\infty}^{\infty} \frac{d\alpha}{2\pi} 
         \exp \lkk -i\alpha(\bar\varphi - \varphi_{R}) 
           - \frac{\beta}{192 \pi^{3/2} R^3} \alpha^2 \rkk \:.
\label{eqn:alpha}
\eea

\noindent
From the above expression, it is evident that, while $\alpha$ runs
from $-\infty$ to $+\infty$, the significant contribution comes from a
finite region

\beq
  -\sqrt{\frac{96\pi^{3/2}R^3}{\beta}} \ltilde 
\alpha \ltilde \sqrt{\frac{96\pi^{3/2}R^3}{\beta}}\equiv \alpha_{\rm m}(R) \:.
\label{eqn:range}
\eeq

\noindent
But we do not perform $\alpha$-integration for the moment.  Instead we
first define the generating functional for the connected Green
functional $W[J]$ as

\beq
  W[J] = \beta^{-1} \ln \lhk
          \frac{1}{\CZ} \int \CD\varphi 
           \exp \lmk -\beta H[\varphi] 
                   + \beta \int d^3 x J(\vect x)\varphi(\vect x)    
                \rmk
             \rhk \:. \label{W}
\eeq

\noindent
Then the probability distribution function is expressed as

\beq
  P[\varphi_{R} = \bar\varphi] =
       \int \frac{d\alpha}{2\pi} \exp \lmk
           \beta W[J] - i \alpha\bar\varphi 
             + \frac{\beta}{2} \int d^3 x 
          \sum_{n\ne0} J(\vect x) \lmk \left. \frac{\delta^2\CL_{E}}
              {\delta\varphi^2} \right|_{n} \rmk^{-1} 
            J(\vect x) \rmk \:.
\label{eqn:probability} 
\eeq

\noindent
Thus we can obtain the probability distribution $P[\varphi_{R} =
\bar\varphi]$ by evaluating $W[J]$ in principle. The generating
functional (\ref{W}) is often expressed perturbatively since the
higher-point connected Green function typically involves higher-powers
of (small) coupling constants. In our case with $J(\vect x) = i\alpha
I(\vect x;R)/\beta$ we can formally write

\bea
  \beta W[J] &=& \sum_{n} \frac{\beta^n}{n!} \int
                  d^3 x_{1} \ldots d^3 x_{n}
                   G_{\rm c}^{(n)}(\vect x_{1} \ldots \vect x_{n})
                    J(\vect x_{1}) \ldots J(\vect x_{n}) \non \\
             &=& \sum_{n} \frac{(i\alpha)^n}{n!} \la (\varphi_{R})^n
                   \ra  \:, 
\label{eqn:gene}
\eea

\noindent
where $\la (\varphi_{R})^n \ra$ is defined by

\beq
  \la (\varphi_{R})^n \ra \equiv \int
        d^3 x_{1} \ldots d^3 x_{n}
                   G_{\rm c}^{(n)}(\vect x_{1} \ldots \vect x_{n})
                    I(\vect x_{1};R) \ldots I(\vect x_{n};R) \:.
\eeq

\noindent
As mentioned above, since the dominant contribution comes from a
finite range of $|\alpha|$, in many cases it is suitable to use only
lower-order terms of (\ref{eqn:gene}) in (\ref{eqn:probability}).

\section{Application to the Electroweak Phase Transition}

\label{sec:application}

\indent 

\begin{figure}[htb]
  \begin{center}
  \leavevmode\psfig{figure=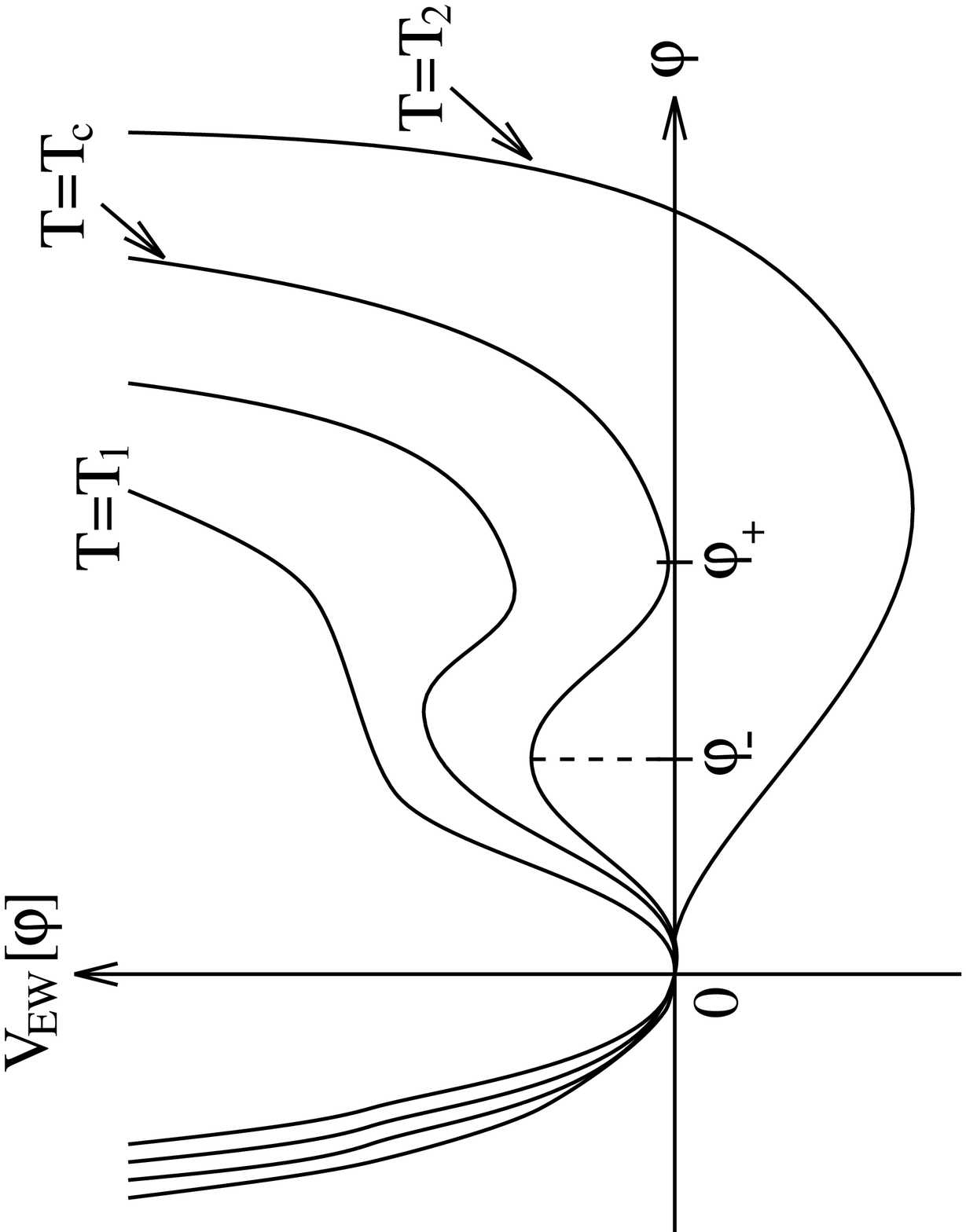,width=10cm}
  \end{center}
  \caption{One-loop improved finite temperature effective potential of
  the Higgs field, $V_{EW}[\varphi]$.}
  \label{fig:potential}
\end{figure}

Having established a generic formalism, we now apply our formalism to
the electroweak phase transition. The one-loop finite-temperature
effective potential of the Higgs field in the minimal standard model
is well approximated by \cite{DLHLL}-\cite{She}

\beq
  V_{EW}[\phi] = D(T^2-T_{2}^2)\phi^2
                     -ET\phi^3
                      +\frac14\lambda_{T}\phi^4 \:, 
\label{eqn:ew}
\eeq

\noindent
where $D=0.169$ and $E=0.00965$. For $M_{H} = 60$GeV, which is the
lower bound allowed experimentally \cite{dat}, we find $\lambda_{T} =
0.035$, $T_{2} = 92.65$GeV, and the critical temperature is given by
$T_{c} = 93.39$GeV. The temperature dependence of the potential is
depicted in Fig.\ \ref{fig:potential}.  At the critical temperature,
the effective potential has two minima, at $\phi=0$ and
$\phi\equiv\phi_+(T_c)=\frac{2ET_c}{\lambda_T}=51.5$GeV, and the
inflection point is equal to $\phi_{\rm inf}=10.9$GeV.

Although the shape of the effective potential exhibits a typical
feature of a first-order phase transition, the potential barrier,
which gets smaller for larger $M_H$, is so shallow that it has been
suggested that the universe was in a mixed state of true and false
vacua already at $T=T_c$ due to subcritical fluctuations \cite{GKW}
\cite{GK} \cite{GRPL}-\cite{SMY} as mentioned in \S
\ref{sec:introduction}.  Furthermore analytical cross-over is observed
for $M_{H} \gtilde 80$GeV in recent results of lattice Monte Carlo
simulations \cite{Ja} \cite{Ru}. On the other hand, by estimating the
amplitude of fluctuations on the so-called correlation length, namely
the curvature scale at the origin of the potential, or on even larger
scales, it has also been claimed subcritical fluctuations is small
enough and the conventional picture of the phase transition works
\cite{DLHLL} \cite{Bet} \cite{ERV}.

The amplitude of thermal fluctuations depends largely on the
coarse-graining scale.  In our recent work \cite{YY}, by solving the
simple Langevin equation, we have shown that although the curvature
scale of the potential or the Compton wavelength is a good measure of
correlation length for a non-selfinteracting massive scalar field in
thermal equilibrium, it is not sensible to take it as the
coarse-graining scale for a field with a more complicated potential
such as the electroweak Higgs field.  We need to determine it from a
fundamental point of view.

In \cite{YY} we have
re-considered derivation of the Langevin equation for the expectation
value of a scalar field at finite temperature originally developed in
\cite{Mori} and \cite{GR} using non-equilibrium quantum field theory.
We have discussed that properties of stochastic noise force in the
Langevin equation have distinct features depending on whether it
arises from interactions with bose fields or from fermi fields, and
shown that the latter is more effective to disturb the system from a
homogeneous configuration.  Hence we expect the correlation properties
of these fermionic noises play an important role in determining the
correlation length of the scalar field.

As shown in \cite{YY}, the spatial correlation function of the
stochastic thermal noise force, $\xi(\vect x,t)$, generated by a
massive fermion $\psi$ through Yukawa coupling, $f\phi\bar\psi\psi$,
is given by

\bea
  \la\,\xi(\vect x,t)\xi(\vect y,t)\,\ra
             &&  \non \\*
             && \hspace{-3.3cm}
                  =\frac{m_{\psi}^4 f^2}{4\pi^4 r^2} \lhk 
                    \lkk\,K_{2}(m_{\psi}r)+2\sum_{n=1}^{\infty}(-1)^n
                          \frac{r^2}{\sqrt{r^2+n^2 \beta^2}}
                           \,K_{2}(m_{\psi}\sqrt{r^2+n^2 \beta^2}\,)
                    \,\rkk^2 \right. \non \\* 
             && \hspace{-3.0cm}    \left.
                   -\lkk\,K_{1}(m_{\psi}r)+2\sum_{n=1}^{\infty}(-1)^n
                          \frac{r}{r^2+n^2 \beta^2}
                           \,K_{1}(m_{\psi}\sqrt{r^2+n^2 \beta^2}\,)
                    \,\rkk^2 \,\rhk\:,~~~r\equiv |\vect x - \vect y| \:,     
\eea

\noindent
where $f$ is a coupling constant and $m_{\psi}$ is the mass of the
coupled fermion $\psi$. It damps exponentially above the inverse mass
scale for $m_{\psi}\beta \gg 1$, and above the scale $\beta$ for
$m_{\psi}\beta \ltilde 1$. In the case with $m_\psi=0$, we find

\bea
  \la\,\xi(\vect x,t)\xi(\vect y,t)\,\ra
            &=& \frac{f^2}{4 \pi^4 r^2}
                  \lkk\, \frac{\pi}{\beta r}
                   \frac{1}{\sinh \lmk \frac{r}{\beta}\pi \rmk} 
                 +\frac{\pi^2}{\beta^2}
                   \frac{\cosh \lmk\frac{r}{\beta}\pi \rmk}   
                        {\sinh^2 \lmk \frac{r}{\beta}\pi \rmk}
                  \,\rkk^2  \:, \\*
            &\simeq& 
                  \frac{f^2}{4\beta^4}
                    \frac{e^{-\frac{2\pi r}{\beta}}}{r^2} 
                       \qquad \mbox{for} \:\: r \gtilde \frac{\beta}{\pi} \:.   
\eea

\noindent
So the typical damping scale is given by $\beta/(2\pi)$. In
both cases the temporal correlation damps exponentially if the
temporal difference becomes larger than $\beta/(2\pi)$.

Therefore we should primarily calculate the PDF taking the
coarse-graining scale equal to this correlation length of the
stochastic noise field acting on $\phi$ because it is the only source
of inhomogeneous evolution of the field.  In practice, however, we
calculate how the PDF changes as we increase the coarse-graining scale
from $\beta/(2\pi)$ to the curvature scale at the origin of the
potential to compare with other work.  Note that it is a good
approximation to treat the noise field as arising from a massless
fermion because, even if the two phases $\phi=0$ and $\phi_+$ are
completely mixing the average mass of the top quark, which is the most
strongly-coupled fermion to the Higgs field and provides the dominant
source of thermal noise, is only about 18 GeV then, much smaller than the
critical temperature.

In order to obtain the proper PDF, we need evaluate the generating
functional $W[J]$. For this purpose, we must calculate the 
finite-temperature three dimensional effective {\it action}, 
which is a formidable task, and so we
approximate it by the standard kinetic term and the effective {\it
  potential}. That is, the one-loop three-dimensional 
effective action is replaced by

\beq
  H[\varphi] = \int d^3 x \lkk\,
                   \frac12 \vect\nabla\varphi(\vect x) \cdot 
                            \vect\nabla\varphi(\vect x) 
                    +V_{\eff}[\varphi]
                             \,\rkk \:. \label{31}
\eeq

\noindent
Strictly speaking, this replacement is justified only if deviation
from homogeneous configuration is sufficiently small. When the field
is in a mixed state between the two minima, the effective potential
should be modified. It is well-known, however, that this modification
reduces the potential barrier \cite{Eff}, which induces more
phase-mixing. Therefore, even if we evaluate the probability
distribution function by using the above approximation and reach the
conclusion that two phases are mixed, the conclusion is consistent.

Below we proceed the discussion by using the above approximation and
adopt $V_{EW}[\varphi]$ as the effective potential
$V_{\eff}[\varphi]$. The alert reader may notice that
$V_{\eff}[\varphi]$ should be that integrated only over heavy modes
for the Higgs self-interaction. But, since the effect of the Higgs
self-interaction is so small that it has already been omitted in
evaluating $V_{EW}[\varphi]$, it makes no change. We fix $T=T_{c}$
in order to investigate whether the universe is in a
homogeneous state of the false vacuum at the onset of the phase
transition. We also set the Higgs mass $M_{H} = 60$GeV. Then
$V_{EW}[\varphi]$ is given by

\beq
  V_{EW}[\varphi] = \frac12 M^2 \varphi^2 - \frac{1}{3!}\mu T_{c}\varphi^3
                    + \frac{1}{4!}\lambda\varphi^4 \:,
\eeq

\noindent
where $M \simeq 0.073T_{c}$ and the dimensionless coupling constants
read $\mu \simeq 0.058$ and $\lambda \simeq 0.21$. Since it is impossible
to evaluate the generating functional $W[J]$ exactly even in the case 
(\ref{31}) is adopted, we  expand
$W[J]$ perturbatively. Also, we neglect the gauge-nonsinglet nature of
the Higgs field for simplicity and consider only its real-neutral
component to treat $\varphi$ as if it was a real singlet field.

\begin{figure}[htb]
  \begin{center}
  \leavevmode\psfig{figure=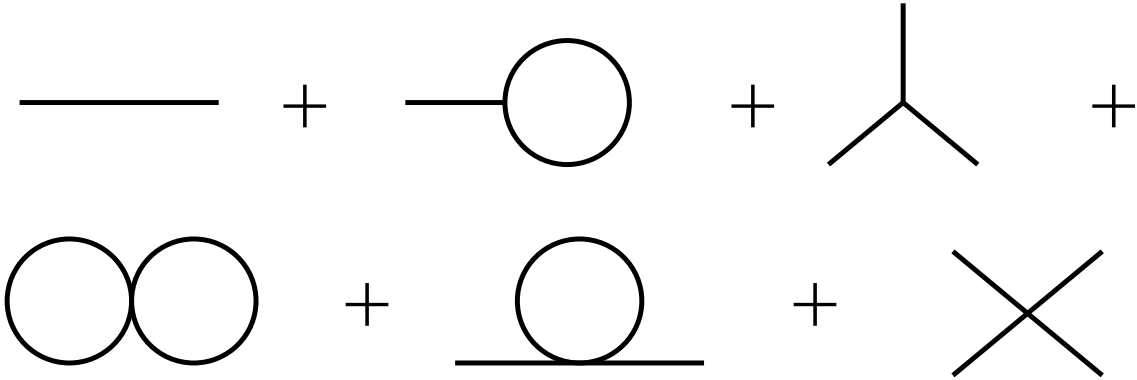,width=15cm}
  \end{center}
  \caption{Contribution to $W[J]$ up to $\CO(\mu, \lambda)$.}  
  \label{fig:green}
\end{figure}

Up to the first order of the coupling constant, $W[J]$ is made up of
the graphs as depicted in Fig.\ref{fig:green}. That is, 

\bea
  \beta W[J] &=& \frac{1}{2} \la J_{x}\Delta_{xy}J_{y} \ra_{xy}
                    \non \\
             &+& \frac{\mu}{3!\beta^3} \lkk\,
                 3\la \Delta_{xx}\Delta_{xa}J_{a} \ra_{xa}
                + \la \Delta_{xa}\Delta_{xb}\Delta_{xc}
                              J_{a}J_{b}J_{c} \ra_{xabc}
                 \,\rkk \non \\
             &-& \frac{\lambda}{4!\beta^3} \lkk\, 
                 3\la \Delta_{xx}^2 \ra_{x} 
                + 6\la \Delta_{xx}\Delta_{xa}\Delta_{xb}
                               J_{a}J_{b} \ra_{xab}
                + \la \Delta_{xa}\Delta_{xb}\Delta_{xc}\Delta_{xd}
                              J_{a}J_{b}J_{c}J_{d} \ra_{xabcd}
                 \,\rkk   \non \\
             &+& ({\rm higher-order~ terms}) \:,
\eea

\noindent
where $\Delta_{xy}$ is given by

\beq
  \Delta_{xy} = \beta \int \frac{d^3 k}{(2\pi)^3}
                  \frac{1}{\vect k^2 + M^2}\, 
                   e^{i\vect k \cdot (\vect x - \vect y)} \:.
\eeq 

\noindent
Here $\la\cdots\ra_{xab\cdots}$ implies integration over $d^3 x d^3 a d^3
b \cdots$, and $\beta = 1 / T_{c}$. Each loop graph can be
renormalized into the definition of $\bar\varphi$, the normalization
of the probability, and the mass, $M$. Therefore, we have only to
evaluate three tree graphs.

First we consider the two-point function. Adding the last term of the
exponent in eq.(\ref{eqn:probability}), it is given by

\bea
  \la (\varphi_{R})^2 \ra_{T} &\equiv&  \la (\varphi_{R})^2 \ra +
      \frac{1}{\beta} \int d^3 x 
          \sum_{n\ne0} I(\vect x;R) \lmk \left. \frac{\delta^2\CL_{E}}
              {\delta\varphi^2} \right|_{n} \rmk^{-1} 
            I(\vect x;R) \non \\  
   && \hspace{-2.3cm} =
       \int d^3 x \int d^3 y I(\vect x;R)
        \frac{1}{\beta} \sum_{n} \int \frac{d^3 k}{(2\pi)^3}
                   \frac{1}{\vect k^2 + M^2 + (2\pi n/\beta)^2}\, 
                    e^{i\vect k \cdot (\vect x - \vect y)} 
             I(\vect x;R) \non \\ 
   && \hspace{-2.3cm} =
       \int d^3 x \int d^3 y 
        \int \frac{d^3 k}{(2\pi)^3}
          \lkk \frac12 + \frac{1}{e^{\beta\omega}-1} \rkk
             \frac{1}{\omega}
        e^{i\vect k \cdot (\vect x - \vect y)} 
         I(\vect x;R)I(\vect y;R) \non \\
   && \hspace{-2.3cm} = 
       \frac{1}{2\pi^2} \int_{0}^{\infty} dk
        \lkk \frac12 + \frac{1}{e^{\beta\omega}-1} \rkk
          \frac{k^2}{\omega} \tilde I(\vect k;R)^2 \:,~~~~
\omega \equiv \sqrt{\vect k^2 + M^2}\:,
\eea    

\noindent
where $\tilde I(\vect k;R)$ is the Fourier transform of the window function,

\bea
  \tilde I(\vect k;R) = \int d^3 x I(\vect x;R)
                               e^{i\vect k\cdot\vect x} 
                      = \exp \lmk -\frac12 R^2 \vect k^2 \rmk \:.
\eea

Similarly, the three- and the four-point functions are given,
respectively, by

\bea
  \la (\varphi_{R})^3 \ra &=& \frac{\mu}{\beta^3}
          \int \frac{d^3 k_{1}}{(2\pi)^3}
            \int \frac{d^3 k_{2}}{(2\pi)^3} 
               f(\vect k_{1})f(\vect k_{2})f(\vect k_{1}+\vect k_{2}) \:,
     \non \\
  \la (\varphi_{R})^4 \ra &=& -\frac{\lambda}{\beta^3}
          \int \frac{d^3 k}{(2\pi)^3} g(\vect k)^2 < 0 \:,
\eea

\noindent
where 

\bea
  f(\vect k) &=& \frac{1}{\vect k^2 + M^2} \tilde I(\vect k;R) \:, \non \\
  g(\vect k) &=& \int \frac{d^3 q}{(2\pi)^3} 
                   f(\vect q)f(\vect k+\vect q) \:.
\eea

In order to confirm that it is sufficient to consider graphs with
lowest-order in coupling constants for the case with the fundamental
coarse-graining scale $R=\beta/(2\pi)$, let us estimate the magnitude
of graphs with $n$-th order of coupling constants on scale
$R\equiv N\beta/(2\pi)$.  For even $n$, the dominant contribution
comes from the graph made of $(n/2-1)$ four-point vertices, which
reads

\bea
  && \hspace{-1.0cm} \frac{1}{n!} \frac{\lambda^{\frac{n}{2}-1}}{\beta^{n-1}}
   \int\frac{d^3 k_{1}}{(2\pi)^3} \cdots 
    \int\frac{d^3 k_{n-1}}{(2\pi)^3}
     f(\vect k_{1}) \cdots f(\vect k_{n-1}) \times \non \\
  && \hspace{2.5cm}
      \frac{1}{(\vect k_{1}+\vect k_{2}+\vect k_{3})^2+M^2} \cdots
       \frac{e^{-(\vect k_{1}+ \cdots +\vect k_{n-1})^2R^2}}
        {(\vect k_{1}+ \cdots +\vect k_{n-1})^2+M^2}
  \non \\
  && \ltilde \frac{1}{n!}
  \frac{\lambda^{\frac{n}{2}-1}}{\beta^{n-1}} 
   \lmk
     \int\frac{d^3 k}{(2\pi)^3} f(\vect k)
   \rmk^{\frac{n}{2}}
   \lmk
     \int\frac{d^3 k}{(2\pi)^3} \frac{f(\vect k)}{\vect k^2+M^2}
   \rmk^{\frac{n}{2}-1} \non \\
  && \ltilde
   \frac{1}{n!} \lmk \frac{\lambda}{0.073} \rmk^{\frac{n}{2}-1}
    2^{-\frac{n}{4}+2}\, 3^{\frac{n}{2}}\, \pi^{-\frac{3n}{2}+1}\,
    N^{n} \:,
\label{eqn:approximation}
\eea

\noindent
where $R$ is assumed to be much smaller than $M^{-1}$. For odd $n$,
the graph with one three-point vertex and $(n-3)/2$ four-point
vertices makes dominant contribution to $W[J]$ and its magnitude can
be estimated similarly. As a result we find, for the case $\alpha$ in
$J$ takes the value $\alpha_{\rm m}(R)$ defined in (\ref{eqn:range}),
the graphs with the lowest order in the coupling constants make
dominant contribution to $W[J]$, provided that the coarse graining
scale satisfies $N \ltilde 10$. If the dominant contribution of the
$\alpha-$integral comes from a narrower region than (\ref{eqn:range}),
the constraint on $N$ can be even weaker.

Thus the PDF of the coarse-grained field $\varphi_R$ up to the first order 
of the coupling constants is given by

\bea
  P[\varphi_{R} = \bar\varphi] &=&
       \int_{-\infty}^{\infty} \frac{d\alpha}{2\pi} \exp \lmk
           - i \alpha\bar\varphi 
           - \frac{\alpha^2}{2} \la (\varphi_{R})^2 \ra_{T} 
           - i \frac{\alpha^3}{3!} \la (\varphi_{R})^3 \ra 
           + \frac{\alpha^4}{4!} \la (\varphi_{R})^4 \ra  \rmk \non \\
  &=& \int_{0}^{\infty} \frac{d\alpha}{\pi} \exp \lmk
           - \frac{\alpha^2}{2} \la (\varphi_{R})^2 \ra_{T}       
           + \frac{\alpha^4}{4!} \la (\varphi_{R})^4 \ra \rmk
                                            \cos \lmk            
             \alpha\bar\varphi 
           + \frac{\alpha^3}{3!} \la (\varphi_{R})^3 \ra \rmk \:.
\eea

\begin{figure}[htb]
  \begin{center}
  \leavevmode\psfig{figure=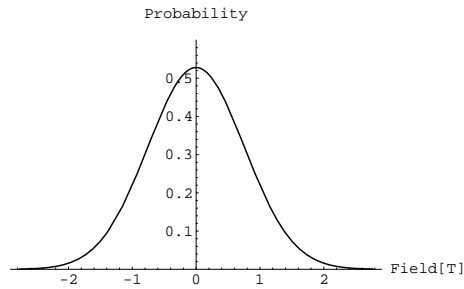,height=6cm}
  \end{center}
  \caption{Probability distribution function for the field
    coarse-grained over $\beta/(2\pi)$. The inflection point, 
    $\varphi_{\rm inf}$ in the potential
    corresponds to $0.117\beta$ and the local maximum, $\varphi_{-}$
    $0.277\beta$.}
  \label{fig:tem}
\end{figure}

The PDF with $R = \beta/(2 \pi)$ is depicted in Fig.\ \ref{fig:tem}.
The probability $\varphi_{R} \geq \varphi_{\rm inf}= 0.117T_c$ is
43.8\% and the root-mean-square (RMS) 
is $0.76T_c$, which is much larger than the
inflection point $\varphi_{\rm inf}$ and the maximum point
$\varphi_{-}= 0.277T_c$ of the potential. This result suggests that
there is non-negligible amount of soaking into the asymmetric phase.
Therefore we can conclude that the electroweak phase transition does
not proceed by the standard bubble nucleation picture.

\begin{figure}[htb]
  \begin{center}
  \leavevmode\psfig{figure=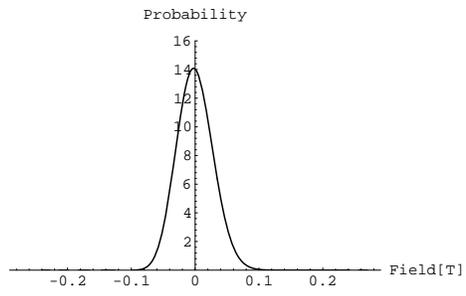,height=6cm}
  \end{center}
  \caption{That over the inverse mass scale.}
  \label{fig:mass}
\end{figure}

For comparison, we show the probability distribution function for the
field coarse-grained over the inverse mass scale $R = 1 / M$ in Fig.\ 
\ref{fig:mass}. In this case, since $R$ is large enough, from the
physical argument, we can infer that the fluctuation becomes small and
approaches the Gaussian configuration. Indeed, the
probability of $\varphi_{R} \geq \varphi_{-}$ is extremely small and
the RMS is $0.028T_c$, much smaller than the inflection point, which
is consistent with the above observation and \cite{Bet}. Table 1 shows
how the probability $\varphi_{R} \geq \varphi_{\rm inf}$ and the
root-mean-square $\sqrt{\la \varphi_R^2 \ra}$ change as a function of
the coarse-graining scale $R$.   Roughly speaking,
if $R \gtilde 40\beta/(2\pi)$, we can neglect the role of these
thermal fluctuations. In the previous work \cite{YY}, the critical
scale is about $R_{L} \simeq 80\beta/(2\pi)$. The relation between the
coarse-graining scale in this paper and that in the previous paper is
not so clear because the former is defined in the spherical Gaussian
window function and the latter corresponds to that defined in the
cubic top-hat window function. But, if the correspondence is decided
by the equality of the volume, that is,

\beq
  R_{L}^3 = \int d^3 x \frac{4\pi r^3}{3} I(\vect 0;R) \:,
\eeq

\noindent
we find $R_{L} \simeq 3.0 R$. Therefore the classical lattice
simulation of \cite{YY} is in reasonable agreement with the present
analysis.

\section{Concluding remark}

\label{sec:remark}

\indent

In this paper, we formulate the probability distribution function for
the coarse-grained fluctuations, which can be evaluated by calculating
perturbatively the connected generating functional $W[J]$. In
application, we apply our formalism to the electroweak phase
transition. We evaluate the probability distribution function of the
field coarse-grained over $2\pi / \beta_{c}$ so that we find that the
thermal fluctuations called subcritical bubbles play a significant
role and that the universe is not homogeneously in the symmetric phase
at the onset of the electroweak phase transition. We also calculate
the probability $P[\varphi_{R} \ge \varphi_{\rm inf}]$ and the
root-mean-square at several coarse-graining scale. The result is
depicted in Table.\ \ref{tab:prob}. We can observe that as the
coarse-graining scale is larger, the probability $P[\varphi_{R} \ge
\varphi_{\rm inf}]$ becomes smaller and also the RMS smaller. Thus,
the coarse-graining scale is very important in investigating the
property of thermal fluctuations. Unless the coarse-graining scale is
specified properly, the discussion becomes meaningless and brings out
confusion.

\begin{table}[tcb]
\caption{Property of thermal fluctuations at several scales} 
\label{tab:prob}
  \begin{center}
     \begin{tabular}{rcc} \hline
        $R~~~~~~~~~~~$ & 
        $P[\varphi_{R} \ge \varphi_{\rm inf}]$ &
        $\sqrt{\la \varphi_R^2\ra}$ \\ \hline
        $~~\beta/(2\pi) \simeq 0.159\beta$  & $44\%$ & $0.76T_c$ \\ 
        $10\beta/(2\pi) \simeq 1.59\beta$ & $22\%$   & $0.15T_c$ \\ 
        $20\beta/(2\pi) \simeq 3.18\beta$ & $12\%$  & $9.8\times10^{-2}T_c$ \\ 
        $30\beta/(2\pi) \simeq 4.77\beta$ & $5.9\%$ & $7.3\times10^{-2}T_c$ \\ 
        $40\beta/(2\pi) \simeq 6.37\beta$ & $2.6\%$ & $5.8\times10^{-2}T_c$ \\ 
        $50\beta/(2\pi) \simeq 7.96\beta$ & $0.99\%$& $4.8\times10^{-2}T_c$ \\ 
        $60\beta/(2\pi) \simeq 9.55\beta$ & $0.32\%$& $4.1\times10^{-2}T_c$ \\ 
        $70\beta/(2\pi) \simeq 11.1\beta$ & $8.4\times10^{-2}\%$ & $3.5\times10
^{-2}T_c$ \\  
        $80\beta/(2\pi) \simeq 12.7\beta$ & $1.8\times10^{-2}\%$ & $3.0\times10
^{-2}T_c$ \\ 
        $1/M \simeq 13.7 \beta$           & $1.6\times10^{-3}\%$ & $2.8\times10
^{-2}T_c$ \\ 
   \hline
     \end{tabular}
  \end{center}
\end{table}

\subsection*{Acknowledgments}
MY would like to thank Dr. T.\ Shiromizu for discussion. MY is
grateful to Professor K. Sato for his continuous encouragement and
Professor M.\ Morikawa for his useful comments. This work was
partially supported by the Japanese Grant in Aid for Scientific
Research Fund of the Ministry of Education, Science, Sports and
Culture Nos.\ 07304033(JY) and 08740202(JY).

\newpage

\end{document}